# Reveal of small alkanes and isomers using calculated core and valence binding energy spectra and total momentum cross sections

Zejin Yang[1,2] and Feng Wang[1*]

[1]*eChemistry Laboratory, Faculty of Life and Social Sciences, Swinburne University of Technology, Hawthorn, Melbourne, Victoria, 3122, Australia.*

[2]*School of Science, Zhejiang University of Technology, Hangzhou, 310023, China;*

[*]Corresponding author: fwang@swin.edu.au

# Abstract

The present study revealed quantum mechanically that the C1s binding energy spectra of the small alkanes (upto six carbons) provide a clear picture of isomeric chemical shift in linear alkanes and branched isomers, whereas the valence binding energy spectra contain more sensitive information regarding the length of the carbon chains. Total momentum cross sections of the alkanes exhibit the information of the chain length as well as constitutional isomers of the small alkanes. The C1s binding energies of small alkanes (including isomers) are position specific and the terminal carbons have the lowest energies. The length of an alkane chain does not apparently affect the C1s energies so that the terminal carbons (289.11 eV) of pentane are almost the same as those of hexane. The valence binding energy spectra of the alkanes are characterized by inner valence and outer valence regions which are separated by an energy gap at approximately 17 eV. The intensities of the total momentum cross sections of the alkanes are proportional to the number of carbons in the alkane chains whereas the shape of the cross sections in small momentum region of $p < 0.5$ a.u. indicates their branched isomers.

# 1. Introduction

Small alkanes are fundamentally important to organic chemistry and to the production of gaseous and liquid fuels, which is composed of ~10% pentanes and hexanes.[1] Alkanes are also prototype molecules to study isomers which possess the same number of atoms but different molecular structures and therefore, exhibit different properties. Although benchmarking studies for small alkanes and their structural isomers have been available for many years,[1] the studies of alkane continue to be an important area of research. For example, a breakthrough of hexane isomer separation study is recently published in Science,[2] accurate carbon C1s binding energy spectra of alkanes were reported[3,4] and gamma-ray Doppler-shift for positron-electron annihilation spectra of small alkanes are also studied.[5-7] As a result, a systematic study of chemical environmental changes of alkane isomers with respect to their binding (vertical ionization) energy spectra and total momentum distributions will help one to obtain insight information to interpret the gamma-ray spectra of alkanes.

Despite of a recent theoretical effort to study small alkanes[4] in order to obtain a comprehensive information for inner shell ionization energies of alkanes measured,[3] to comprehensively understand the electronic structures of alkanes needs information of their orbital energies as well as the orbital shape, *i.e.*, the wavefunctions of the molecules under study. Binding energy spectra of molecules are associated with their molecular orbitals where the electrons are ionized.

In his well-known remark some fifty years ago, Coulson[8] stated that "the most important clue seems to me to be the recognition that the energy is not the only goodness of a wavefunction. In the past we have been preoccupied with energy." Although the importance of wavefunctions has been recognised, a quantitative measurement of wavefunctions (orbitals) of a molecule is still a difficult task, which restricts the development in this direction. Until now, electron momentum spectroscopy (EMS)[9,10] is still the dominant experimental technique available for such measurements, in which molecular orbitals (orbital cross sections) can be measured quantitatively in momentum space. The dual space analysis (DSA)[11] is based on EMS which has been proven an excellent method to assess the quality of wavefunctions produced from a quantum mechanical model.[11]

Electron momentum spectroscopy is a powerful tool when combined with high resolution synchrotron sourced X-ray photoemission and absorption spectroscopy, especially for the isomers and conformations due to subtle differences.[12-15] However, almost all such studies focused on the valence space of the molecules such as the highest occupied molecular orbitals (HOMOs). Core space information were much rare,[3,4] as many properties of molecules are determined by frontier molecular orbitals in valence space. Moreover, even less known of the molecular structural effects in momentum space.[16-19] In addition, systematic studies of isomerization effects of even small alkanes with respect to their binding energy spectra are rare.[3,4,11] As a result, the valence and core (C1s) binding energy spectra, and the orbital (site) dependence on the isomerization of the alkanes, deserve a more detailed study.

Recent studies of the Doppler-shift of gamma-ray spectra of small alkanes[6,7,20,21] in positron-electron annihilation processes indicated that positrophilic electrons[6] of a molecule are dominated by inner valence electrons on the lowest occupied valence orbital (LOVO),[7] rather than the HOMO in electron chemistry. As gamma-ray spectra of molecules highly depend on the total momentum distributions (cross sections) of molecules,[20,21] it is desirable for a systematic study of isomeric effects on the binding energy spectra and on the total momentum distributions of small alkanes.

In the present study, detailed core (C1s) and valence binding energy spectra of the isomers of small alkanes containing carbon atoms up to six (methane-hexane) are calculated quantum mechanically, and their C1s chemical shifts with respect to their linear alkane isomers containing the same number of carbons are presented, together with the valence binding energy spectra of these alkanes. Total orbital momentum distributions of the alkanes (n=1-6) and their isomers (n=4-6) are obtained from the Fourier transform of the orbitals calculated in coordinate space.[11] The present study analyses the binding energy spectral information to understand the structural impact on the properties of the alkanes and their isomers systematically.

## 2. Methods and computational details

Geometries of small saturated alkanes (n≤6) and their isomers (n=4-6) in the ground electronic states are optimized systematically using the density functional theory (DFT) based B3LYP/aug-cc-pVTZ model. Vertical ionization potentials of the alkanes are obtained using the OVGF/TZVP model (valence), the SAOP/et-pVQZ[22] (valence) and the LB94/et-pVQZ[16,23] models (core). The B3LYP and OVGF calculations are produced using the Gaussian 03[24] computational chemistry package (with Gaussian basis sets), whereas the SAOP/et-pVQZ and LB94/et-pVQZ calculations are generated by the ADF[25] computational chemistry package (with Slater basis sets).

The valence binding (*i.e.*, vertical ionization) energy spectra of the alkanes are simulated using a Gaussian shape function with full width at half maximum (FWHM) of 0.136 eV, which represents a higher resolution than currently available synchrotron sourced x-ray photoemission spectroscopy with a typical resolution of 0.40 eV for most alkanes,[26] usually varying from 0.27 eV to 0.86 eV.[27]

Molecular orbitals (MO) obtained in coordinate space (based on the B3LYP/TZVP model) were directly mapped into momentum space as theoretical momentum distributions (TMD). The overlap between the target-ion is the one electron Dyson orbital,[8]

$$\sigma \propto \int d\Omega |\phi_j(\vec{p})|^2.$$

Here, $\vec{p}$ is the momentum of the target electron at the instant of ionisation. As a result, the total momentum distributions of the alkanes are the superposition of the orbital momentum distributions of all orbital momentum distributions of the molecule.

## 3. Results and discussion

### 3.1 Geometry and properties of the alkane isomers

The calculated properties of the small alkanes and their isomers obtained using the DFT-B3LYP/aug-cc-pVTZ model are summarised in Table 1. Alkanes are inert compounds whose permanent dipole moment is very small (< 0.15 Debye) or zero, depending on their point group symmetry. For linear alkanes with carbon atoms up to six, these with only odd number of carbons, *i.e.*, propane (μ=0.09 Debye) and pentane (μ=0.09 Debye) have small permanent dipole moment. The branched alkane isomers may possess non-zero permanent dipole moment and the lower the point group symmetry, the larger the permanent dipole moment. For example, the point group symmetry of 2-methypentane (*i.e.*, isohexane) and 3-methylpentane is $C_1$ and $C_s$, respectively, their permanent dipole moment is 0.13 Debye and 0.10 Debye, respectively.

As the present study concerns C1s binding energy spectra of the alkanes, nuclear repulsion energies (NREs) of the alkanes are calculated. Figure 1 presents the relative NRE diagram of the alkanes. The optimized structures of the alkanes are implemented in a recently developed three-dimensional (3D) portable document format (PDF).[28] Double click on the structures in Figure 1 (in a computer) will activate the embedded interactive 3D structures of the isomers in 3D space. For linear alkanes (first column of this figure) from methane to n-hexane, when the carbon chain extends with an additional methyl group (-CH$_3$) attached at a terminal, the NRE increases significantly. For example, the NRE increases approximately 30 E$_h$ from methane to ethane which is almost doubled than the NRE increase from pentane to hexane. Among isomers (n ≥ 4) containing the same number of carbon atoms, it has been well documented that only small total energy differences are presented.[1] However, as shown in Figure 1, that the NREs are significantly different. For example, the linear n-hexane possesses the lowest NRE among its five structural isomers, whereas 2,2-dimethylbutane exhibits the largest NRE in Figure 1.

**3.2 Inner shell binding energy spectra**

Inner shell (core C1s) ionization potentials of a molecule are a property known to be very sensitive to the chemical environment,[3,4,16-19,29] as inner shell spectra provide atomic site specific information and their chemical environment of a molecule. The present study proceeds to study the C1s ionization spectra of the alkanes and their isomers. Figure 2 reports the calculated C1s binding energy (vertical ionization) spectra of the linear alkanes upto hexane. With a small shift,[16-19] the calculated C1s energies are in good agreement with recent theoretical and experimental studies of C1s photoelectron spectroscopic (PES) study of alkane.[3,4] It is noted that the experimentally measured C1s energy (adiabatic) of methane is given by 290.703 eV at 298 K,[1] whereas the present calculated energy is vertical ionization energies at 0 K.

The C1s energies of the linear alkanes in Figure 2 exhibit certain symmetry with respect to the symmetric centre of the molecule---for alkanes with odd numbers of carbons, the symmetry centre locates on the centre carbon atom such as $C_{(3)}$ in pentane; whereas for alkanes with even numbers of carbons, the symmetric centre positions at the mid-point of the central C-C bond of the alkane such as $C_{(2)}-C_{(3)}$ of butane. The C1s energies of a linear alkane exhibit a "Λ" shape such that the terminal carbons, $C_{(1)}$ and $C_{(t)}$, possess the lowest C1s binding energies and the centre carbon atom(s) have the highest C1s energy. For example, in normal hexane, which possesses high point group symmetry of $C_{2h}$ and therefore has three pairs of doubly degenerated C1s binding energies, the C1s binding energies of terminal (primary) carbons, $C_{(1)}$ and $C_{(6)}$, have the lowest energy at 289.11 eV, centre carbon atoms of $C_{(3)}$ and $C_{(4)}$ at 289.34 eV and the C1s energies of other secondary carbons, *i.e.*, $C_{(2)}$ and $C_{(5)}$ are given by 289.28 eV, in agreement with available results in literature[1,3].

The corresponding C1s bonding energies of the linear alkanes remain almost unchanged from propane to hexane, regardless the length of the carbon chains. For example, as shown in Figure 2, the C1s energies for the terminal carbons, $C_{(1)}$ and $C_{(t)}$ are within 289.11-289.12 eV from propane to hexane, whereas the C1s energies of $C_{(2)}$ and $C_{(t-1)}$ of the alkanes increase to 289.28 - 289.35 eV in these alkanes, and the C1s of the middle carbon atom(s), $C_{(3)}$ (and $C_{(4)}$) of the alkanes increase again to 289.34 – 289.35 eV from butane to hexane. The C1s binding energy spectra from propane to hexane split into two major clusters: the spectral peak cluster with lower energy which is assigned to the terminal carbons, $C_{(1)}$ and $C_{(t)}$, and the spectral peak cluster with higher binding energy which is formed by all the other carbon atoms except for terminal carbons. For example, in the C1s binding energy spectrum of n-pentane, the lower energy peak at 289.12 eV is given by C1s of $C_{(1)}$ and $C_{(5)}$, whereas the peak cluster with a larger energy centred at ca. 289.30 eV is given by the C1s energies of $C_{(2)}$, $C_{(4)}$ and $C_{(3)}$. The calculated C1s energies of the alkanes are given in Table S1 (in Supplementary Material Document No.________).

The C1s binding energies of the alkane isomers exhibit more significant chemical shift than increase of the linear carbon chain. Figure 3(a) and (b) present the C1s binding energy spectra of (a) butane isomers (n-butane and isobutne) and pentane isomers (n-pentane, isopentane and neopentane), and (b) the C1s energies of the five hexane isomers. In Figure 3(a), the C1s binding energy spectra of butane and isobutane are indeed significantly different: in isobutane, the C1s energy of three equivalent terminal carbons, $C_{(1)}$, $C_{(2)}$ and $C_{(4)}$ are well separated from the C1s energy of the node carbon $C_{(2)}$ which possesses a apparently larger C1s energy. The chemical shift of isobutane is 0.59 eV which is larger than double of the chemical shift in n-butane of 0.23 eV. Similarly, the chemical shifts of isopentane with three major C1s clusters are 0.25 eV and 0.54 eV with respect to the terminal carbon atoms, whereas the chemical shift of the two major C1s peaks of neopentane is as large as 0.88 eV, which the C1s energy of the node carbon, $C_{(2)}$, chemically shifts further away from the terminal carbons.

The trend of the C1s binding energy chemical shift in linear alkane and branched butane and pentane in Figure 3(a) is clearly supported by the C1s binding energy chemical shift of hexane and its isomers, as given in Figure 3(b). Although all the isomers have the same carbons and hydrogen of $C_6H_{14}$, their C1s energies are very different and lack of similarities. The spectra can only be interpreted based on their molecular structures, i.e., structure sensitive. The C1s energies of the terminal carbons locate at lower energy end of the spectra as seen in their linear counterparts. However, the branched isomers produce more than two terminal carbons which all move to the lower end of the binding energy spectra, in the price of a large chemical shift of the C1s energies of the node carbons to a higher energy position in the spectra. For example, 2-methylpentane produces clusters of C1s energies: three terminal carbons, $C_{(1)}$, $C_{(6)}$ and $C_{(5)}$, and a node carbon $C_{(2)}$ at the higher energy end of the spectra and the middle carbons, $C_{(3)}$ and $C_{(4)}$. This C1s energy chemical shift between the terminal and node carbons are more significant in 2,2-dimethylbutane in this same figure. The largest chemical shift is as large as 0.86 eV.

The significant C1s binding energy of alkane isomers is reflected by their NRE given in Figure 1. The isomer related NRE changes can be as large as the energy changes from methane to ethane. For example, the NRE change of methane to ethane is given by 28.87 $E_h$ whereas the NRE change between hexane and 2,2-dimethylbutane is given by 21.64 $E_h$. Similarly significant C1s and N1s binding energy changes have been calculated among tautomers of DNA bases.[16-18,30] As a result, for a comprehensive understanding of electronic structures of compounds, in addition to valence space, one also needs to understand the information of core shell and many computational cost effective frozen core calculations may lead to inaccurate information and sometimes can be even leading information.

**3.3 Valence binding energy spectra**

Isomers of alkanes result in reconfiguration of the entire electronic space of the molecules and thus, different valence binding energies.[18] Small alkanes are prototypes of the frontier orbital theory in organic chemistry, as valence electrons play an important role in their chemical bonding and chemical reactions. Table 2 reports the first ionization potential (IP), i.e., the vertical ionization energy of the highest occupied molecular orbitals (HOMOs) of the alkanes and their isomers, calculated using the OVGF/TZVP model and the SAOP/et-pVQZ model, respectively. Although small model dependent discrepancies between the calculated first IPs exist, the agreement with available measurements and other theoretical results are very good. Note that the error bars of the available results (both experimental measurements and theoretical calculations) are quite large. For example, the first IP of n-butane was given by 12.36 eV[31] and 11.09 eV,[32] which differ by 1.27 eV whereas in the present calculated IPs of the same molecule is given by 12.10 eV by the SAOP/et-pVQZ model and 11.41 eV by the OVGF/TZVP model, which differ by 0.70 eV. However, in general, the first IPs of the alkanes which are calculated using the SAOP/et-pVQZ model are consistently larger that the IPs obtained from the OVGF/TZVP model. The OVGF model usually produces more accurate first IPs for a number of compounds,[33] but the computational costs hardly warrant such accurate calculations for larges compounds beyond hexane. The computational advantages of the SAOP/et-pVQZ model make it very attractive on larger alkane molecules.[33]

The first IPs of the alkanes in Table 2 indicate a general trends of IP energy reduction as the number of carbon atoms in the alkanes increases. For example, the first IPs of methane, ethane, propane, butane, pentane and hexane are calculated as 13.95 eV, 12.63 eV, 12.23 eV, 12.10 eV, 11.96 eV and 11.79 eV, respectively, using the SAOP/et-pVQZ model, which is consistent with the available measurements in the same table. On the other hand, the first IPs of the isomers with the same number of carbon atoms are very close in energy. The hexane isomers possess very close the first IPs with the energy variations of the hexane isomers are within 0.09 eV.

For small alkanes and their isomers up to six carbons, the binding energy changes in fact depend on the additions or positions of the methyl groups. In order to understand methyl caused changes within the frame of the molecular layout, Figure 4 displays the electron density of the HOMOs. As seen in this figure that the HOMOs of small alkanes exhibit a maximum number of nodal planes in particular in the linear alkanes. The nodal planes may contribute to the energy increase of the linear alkanes with respect to their branched counterparts such as isobutane and neopentane which possess fewer nodal planes. In addition, the HOMO of n-butane is $2b_g$ given by the present model. As pointed out before,[11] the orbital energy differences of the valence orbitals for n-butane are very small so that the valence space configuration of normal butane may depend on the model.

Figure 5 reports the valence binding energy spectra of the linear alkanes (n=1-6). The corresponding electronic configurations in the ground states of the alkanes obtained from our B3LYP/aug-cc-pVTZ model are given in Table S2 (See Supplementary Material Document No.________). The valence binding energy spectra show an interesting pattern in the linear alkanes. Addition of a methyl group into the chain results in the next alkane and brings in four more doubly occupied orbitals to the larger alkane. If the complete electronic configuration space of a linear alkane is grouped as core shell, inner valence shell and outer valence shell, the picture is clear when a methyl group is added in: one orbital locates in the core shell, one orbital lies in the inner valence and two orbitals fall in the outer valence shell. The inner and outer valence regions of the valence binding energy spectra of the linear alkanes are clearly separated by a line at ca. 17 eV, which is also seen from the valence binding energy spectra of nucleosides at the energy position at ca. 26 eV.[28] The number of spectral peaks in the inner valence shell of the alkanes is the same as the number of carbon atoms in the alkanes. However, the inner valence binding energies of the linear alkanes spread both directions with respect to the centre of the energy band at ca. 22 eV ($1a_1$ electrons in methane). In the outer valence space, whereas the number of orbitals and therefore spectral peaks (including the degenerated ones) is given by 2n+1 where n is the number of the carbon atoms in the chain.

Similar to the inner valence shell binding energy spectra, the spectral peaks in the outer valence spectra increase by two and also spread both directions with respect to IP=13.95 eV of the HOMO ($1t_1$) of methane. As a result, the HOMO-LUMO energy gaps of the alkanes decrease as the carbon chain grows as shown in Figure 5 (note that the "spectral peaks" of LUMOs do not exist and the positions are marked for references). For example, the HOMO-LUMO gap of methane is given by 9.91 eV whereas this energy gap of n-hexane is calculated as 7.63 eV, a reduction of 2.28 eV. As both the inner valence and outer valence regions spread when the carbon chain grows (*i.e.*, n increases), the energy gap separating the inner and outer valence shells reduces so that no apparent such energy gap is observed for larger alkanes. For example, the binding energy gap between the inner and outer valence spectral regions of ethane is ca. 8 eV whereas this energy gap reduces to ca 2 eV in n-pentane.

Figure 6 reports the calculated valence binding energy spectra of isomers of hexane. The valence binding energy spectra of the hexane isomers are more or less similar: the HOMO-LUMO energy gap of the isomers is ca. 6.8 eV, whereas the inner valence and outer valence regions are well separated by an energy gap of ca. 1eV. Both energy gaps decrease as more branches appear in the isomer. While the outer valence orbitals cluster tightly within an energy range of approximately 4 eV for all these isomers, the spectral peaks in the inner valence energy region of the linear hexane are closer in energy to one another, whereas these peaks of 2,2-dimethylbutane spread into a larger energy region.

### 3.4 Total momentum cross sections as probe for alkane isomers

The calculated binding energy spectra of alkanes in coordination space reflect changes in the alkanes and their isomers. To probe orbital shape (wavefunctions) dependency of the isomer chemical environment differences, we employ the total momentum cross sections of the alkanes, i.e., dual space analysis (DSA)[11]. In their well-known measurements of gamma-ray spectra of alkanes, Iwata *et al*[5] revealed that the Doppler shift of the gamma-ray spectra does not change significantly among the alkanes (upto dodecane), suggesting certain similarities in their total momentum cross sections.[20,34] As a result, the present study further calculates the total valence momentum cross sections of the alkanes and their isomers.

Figure 7 reports the total valence momentum distributions of the small alkanes and their isomers. The total valence momentum cross sections of the small alkanes exhibit half-bell shaped cross sections, indicating strong 2$s$-electron dominance of the cross sections.[11] The relative intensities of the total cross sections (*i.e.* the areas under the curves) are proportional to the number of valence electrons. For example, the total cross section of methane is significantly smaller than that of hexane. The cross sections in this figure also reflect the isomer dependent characters of branched alkanes, which only differ by the relative intensities of the cross sections in small momentum region. For example, the total cross sections of butane (n=4), pentane (n=5) and hexane (n=4) split in the region of $p < 0.5$ a.u., which corresponds to long range region in coordinate space.[10,11] A further analysis reveals that n-alkanes exhibit the maximum cross sections (the solid cross sections in Figure 7) at $p = 0$ a.u., such as n-butane, n-pentane and n-hexane. The total cross sections of the monobranched alkane(s) such as isopentane come next, and finally the dibranched isomer(s) such as neopetane possesses the smallest total cross sections at p = 0.0 a.u. The insert panel demonstrates the apparent shape and intensity differences between the normal hexane and its isomers, in particular, with respect to dibranched 2,2-dimethylbutane.

## Conclusions

Electron vertical ionization (binding) energy spectra in both core and valence space and total valence momentum distributions of small alkanes upto six carbons and their isomers are studied systematically. Appropriate models such as density functional theory (DFT) based models and outer valence Green function (OGVF) based model are employed to calculate the binding energy spectra and their total momentum cross sections for the alkanes and their isomers. The present study discovered that the alkane isomeric changes such as monobranched and dibranched isomers can be well revealed by core sensitive C1s binding energies of the small alkanes. The C1s energies exhibit a "Λ" shape, i.e., C1s (terminal – primary – carbons) < C1s (secondary carbons) < C1s (node – tertiary – carbons). The length of carbon chains does not apparently affect the C1s energies and the C1s energies of the terminal carbons of pentane are almost the same as those of hexane.

The number of carbon atoms in the alkanes or the length of the carbon chains of the alkane can be more sensitively revealed by the valence binding energy spectra. Properties such as the first IPs, HOMO-LUMO energy gaps and the energy gaps between outer valence and inner valence regions are all sensitive to the number of carbon atoms in the alkanes as well as the branches. Moreover, the valence binding energy spectra of the alkanes are characterized by inner valence and outer valence regions separated at ca. 17 eV. The total momentum cross sections of the alkanes are sensitive to both the number of carbon atoms as well as the isomers of the small alkanes. The intensities of the total cross sections of the alkane are proportional to the number of carbon chains whereas the splitting in the small momentum region of p <0.5 a.u. indicates the branched isomers.

**ACKNOWLEDGMENTS**

This work is supported by the Australian Research Council (ARC) under the Discovery Project (DP) Scheme and an award under the Merit Allocation Scheme on the National Computational Infrastructure Facility at the ANU. Swinburne University Supercomputing Facilities should also be acknowledged. FW wish to thank Prof. K. Brove for useful discussions. One of the authors, ZY, thanks a Postgraduate Research Scholarship to Australia from Ministry of Education (China), Swinburne University of Technology (SUT) for hospitality and the financial support from the National Natural Science Foundation of China (Grant No: 11104247).

See Supplementary Material Document No.__________ for C1s binding energies (vertical) for the small alkanes (S1) and valence configurations of the alkanes and their isomers in ground electronic states (S2).

*Yang and Wang*

**Table** captions:

**Table 1.** Properties of small alkanes and their isomers based on the optimized geometries using the B3LYP/aug-cc-pVTZ model.

**Table 2.** Comparison of the first vertical ionization potentials of small alkanes using the SAOP/et-PVQZ model and OVGF/aug-cc-pVTZ model, with available results (eV).

**Figure** captions:

**Figure 1.** Relative nuclear repulsion energies (NREs) of small alkanes and their isomers (the optimized stable structures of the alkanes are shown in 3-D PDF format---double click on a structure when view the pdf file on a computer to activate).

**Figure 2.** Calculated carbon C1s binding energy spectra of linear alkanes (methane to hexane) using the LB94/et-PVQZ model.

**Figure 3 (a)** Calculated carbon C1s binding energy spectra of isomers for butane (n=4) and pentane (n=5) using the LB94/et-PVQZ model.

**Figure 3 (b)** Calculated carbon C1s binding energy spectra of isomers for hexane (n=6) using the LB94/et-PVQZ model.

**Figure 4.** The highest occupied molecular orbitals (HOMOs) of the small alkanes.

**Figure 5.** Calculated valence binding energy spectra of small linear alkanes (n=1-6) using the SAOP/et-PVQZ model.

**Figure 6.** Calculated valence binding energy spectra of hexane isomers using the SAOP/et-PVQZ model.

**Figure 7.** Calculated total momentum cross sections of small alkanes and their isomers.

**Table 1.**

| Mol. | Name | Sym. | State | $\mu$ / Debye | $\langle R^2 \rangle$ (a.u.) | Structure |
|---|---|---|---|---|---|---|
| $CH_4$ | methane | $T_d$ | $X^1A_1$ | 0.0 | 35.85 | $CH_4$ |
| $C_2H_6$ | ethane | $D_{3d}$ | $X^1A_{1g}$ | 0.0 | 110.05 | $CH_3$-$CH_3$ |
| $C_3H_8$ | propane | $C_{2v}$ | $X^1A_1$ | 0.09 | 227.64 | 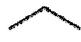 |
| $C_4H_{10}$ | nbutane | $C_{2h}$ | $X^1A_g$ | 0.0 | 426.62 | 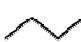 |
| | isobutane | $C_{3v}$ | $X^1A_1$ | 0.14 | 355.31 | 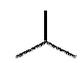 |
| $C_5H_{12}$ | npentane | $C_{2v}$ | $X^1A_1$ | 0.09 | 720.69 | 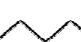 |
| | isopentane | $C_1$ | $X^1A$ | 0.10 | 565.89 | 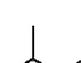 |
| | neopentane | $T_d$ | $X^1A_1$ | 0.0 | 486.43 | 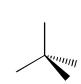 |
| $C_6H_{14}$ | nhexane | $C_{2h}$ | $X^1A_g$ | 0.0 | 1141.18 | 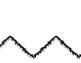 |
| | 2-methylpentane (isohexane) | $C_1$ | $X^1A$ | 0.13 | 906.19 | 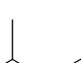 |
| | 3-methylpentane | $C_s$ | $X^1A'$ | 0.10 | 854.89 | 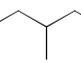 |
| | 2,3-dimethylbutane | $C_{2h}$ | $X^1A_g$ | 0.0 | 746.84 | 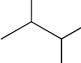 |
| | 2,2-dimethylbutane (neohexane) | $C_s$ | $X^1A'$ | 0.06 | 704.29 | 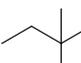 |

**Table 2.**

|  | SAOP et-pVQZ | OVGF TZVP | Exp. | Other Theory |
|---|---|---|---|---|
| methane | 13.95 | 14.10 | 14.00[35,36] | 14.3;[37]14.30[38] |
| ethane | 12.63 | 12.43 | 12.10;[35]12.50[39] | 13.21;[40]13.2[39] |
| propane | 12.23 | 11.82 | 13.22;[33]11.50[35] | 12.07,12.71[41] |
| nbutane | 12.10 | 11.41 | 12.36;[42]11.09[43] | 12.09;[44]12.42[45] |
| isobutane | 12.09 | 11.48 | 12.47;[33]12.51[42] | 12.44;[40]12.46[46] |
| npentane | 11.96 | 11.10 | 10.37,14.84[33] | 12.24;[33]11.97[40] |
| isopentane | 11.89 | 11.05 | 10.32,13.99[33] | 12.12,11.86[40] |
| neopentane | 12.05 | 11.24 | 10.40,12.40[33] | 12.31;[33]12.14[40] |
| nhexane | 11.79 | 10.85 | 10.27;[33]10.43[47] | 10.32;[37]11.20[44] |
| isohexane | 11.79 | 10.83 |  |  |
| 3-methylpentane | 11.76 | 10.83 |  |  |
| 2,3-dimethylbutane | 11.70 | 10.81 |  |  |
| 2,2-dimethylbutane | 11.71 | 10.84 |  |  |

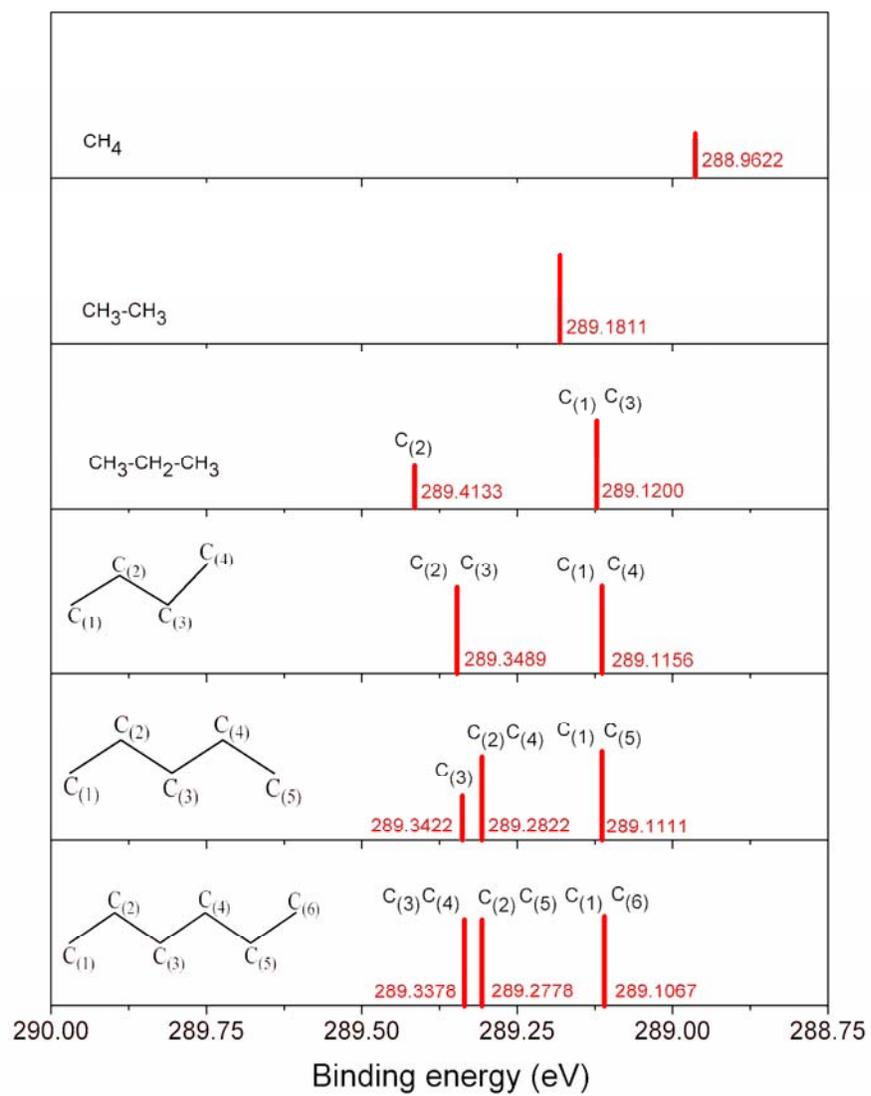

**Figure 2.**

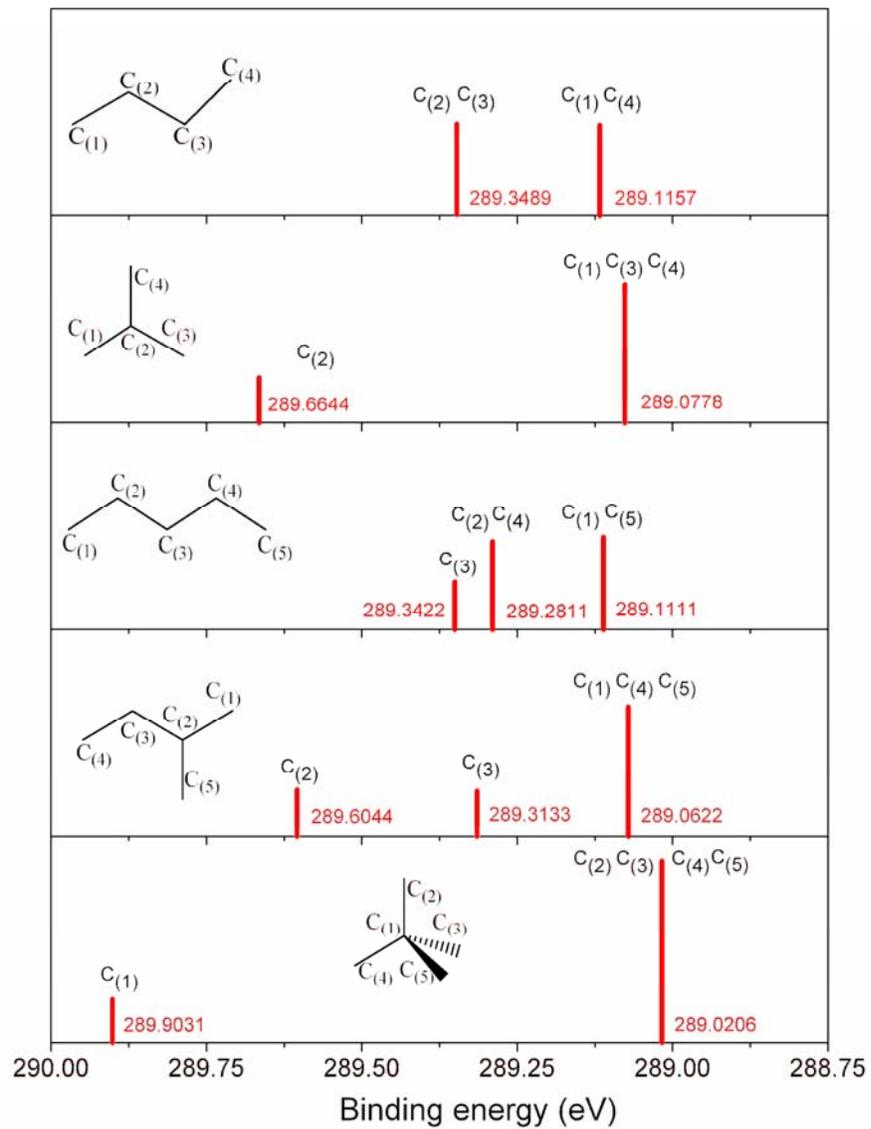

**Figure 3 (a)**

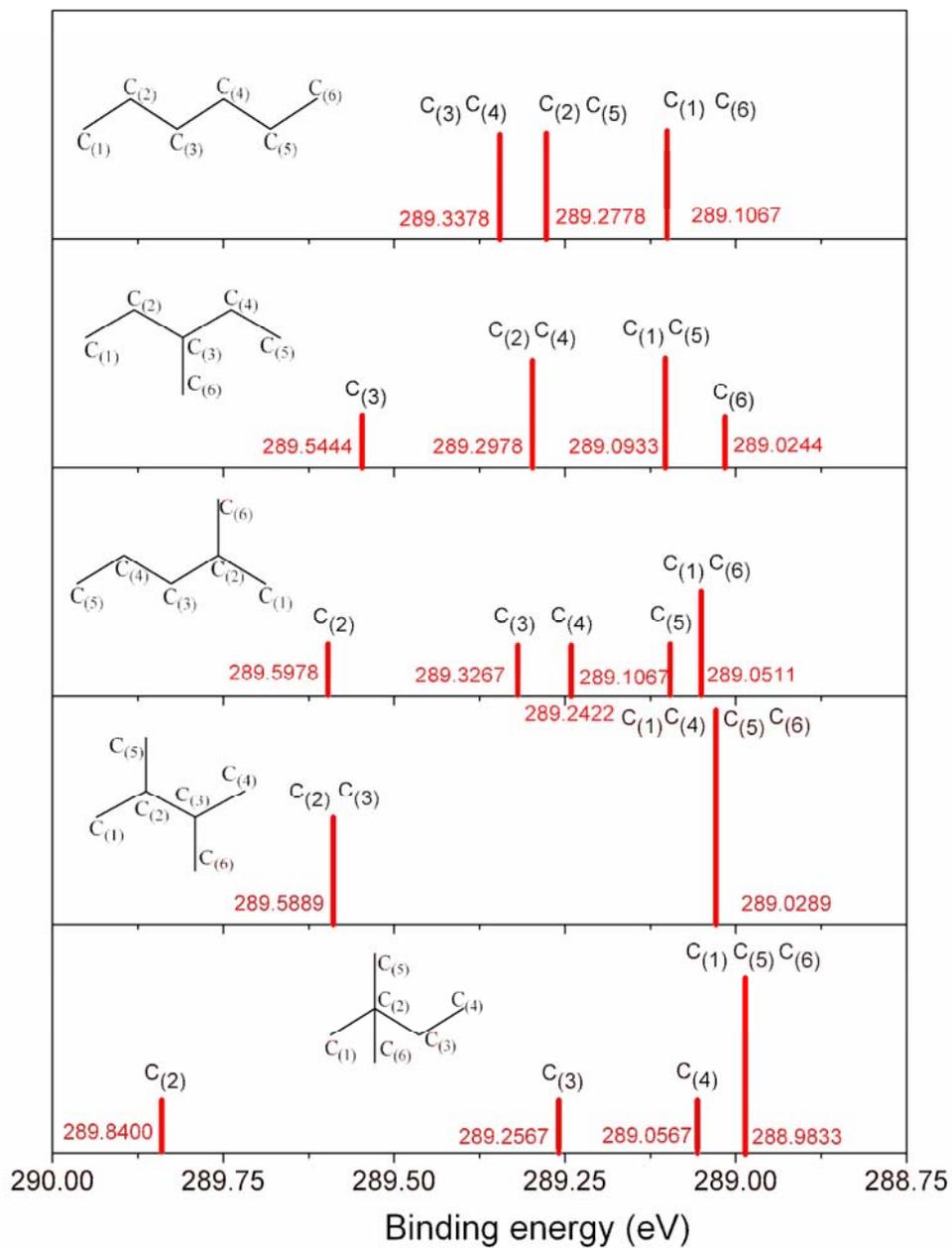

**Figure 3 (b)**

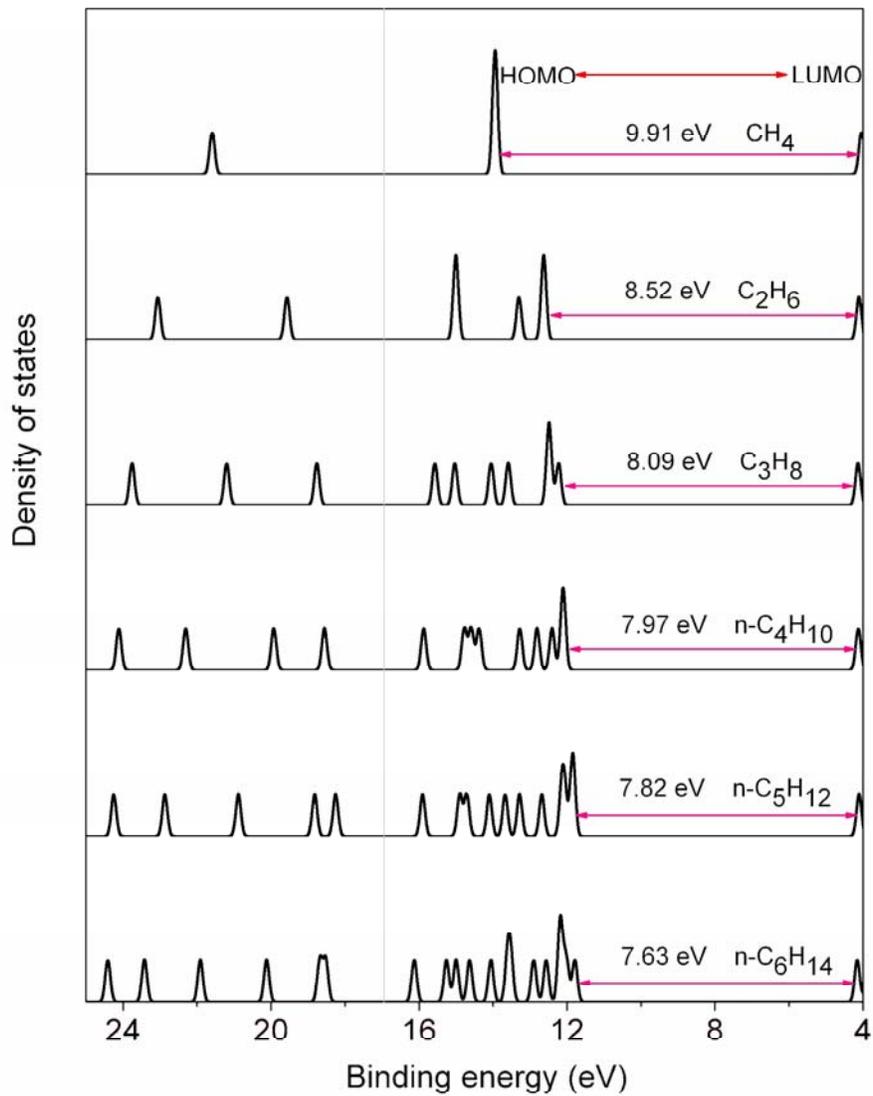

**Figure 5.**

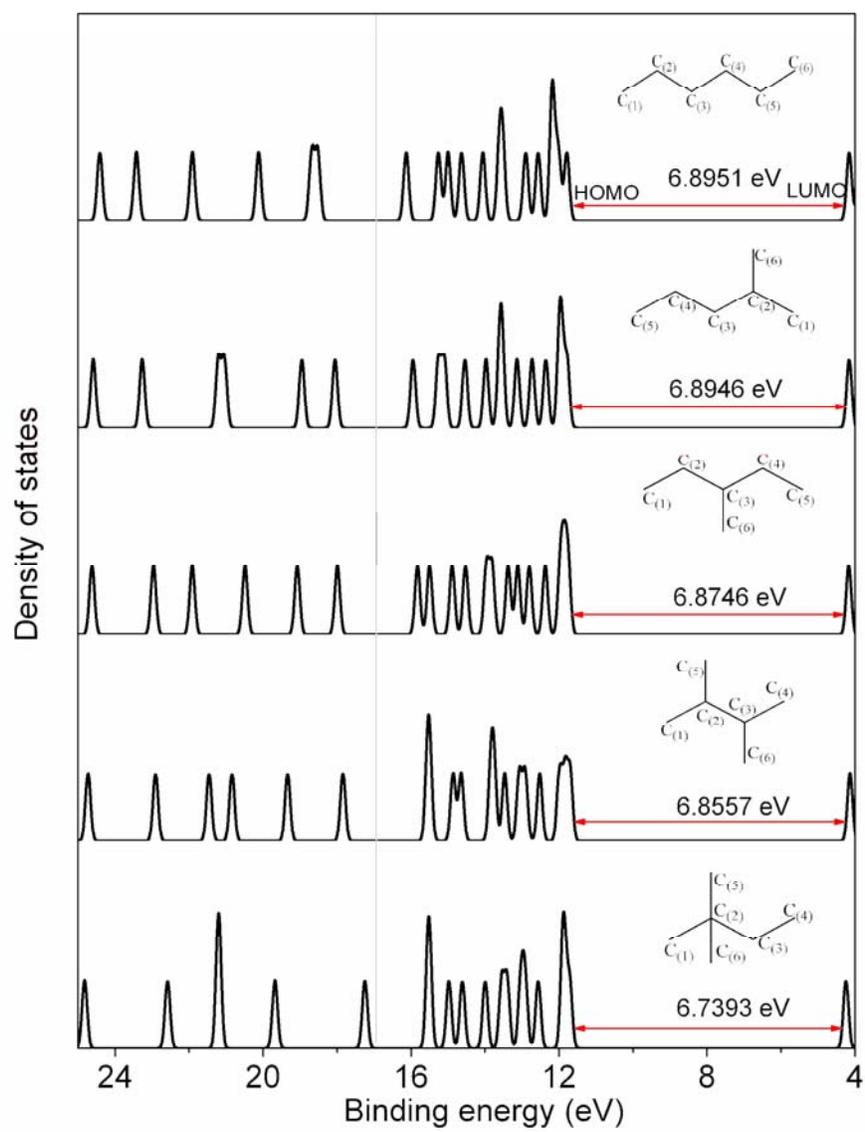

**Figure 6.**

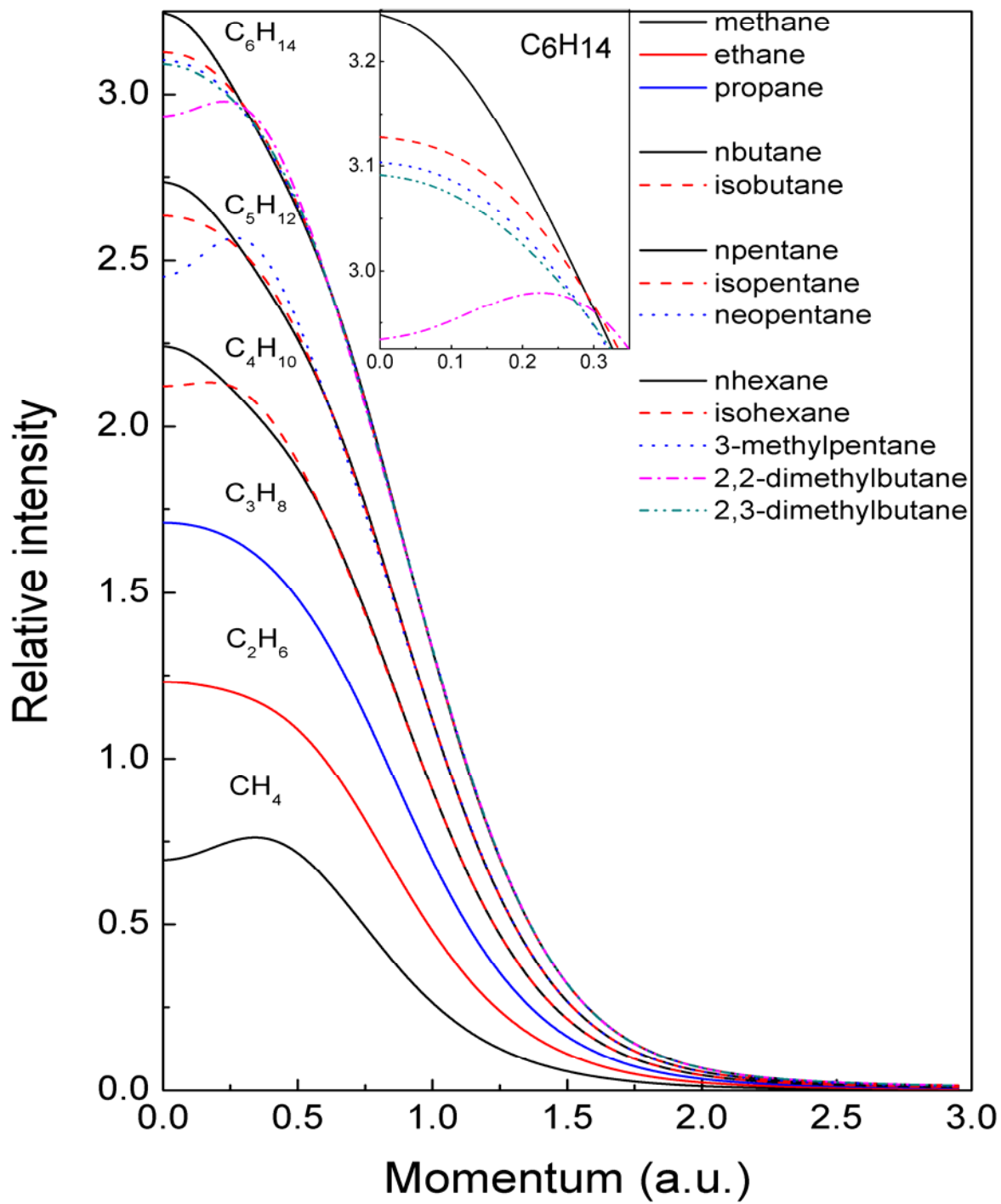

**Figure 7.**

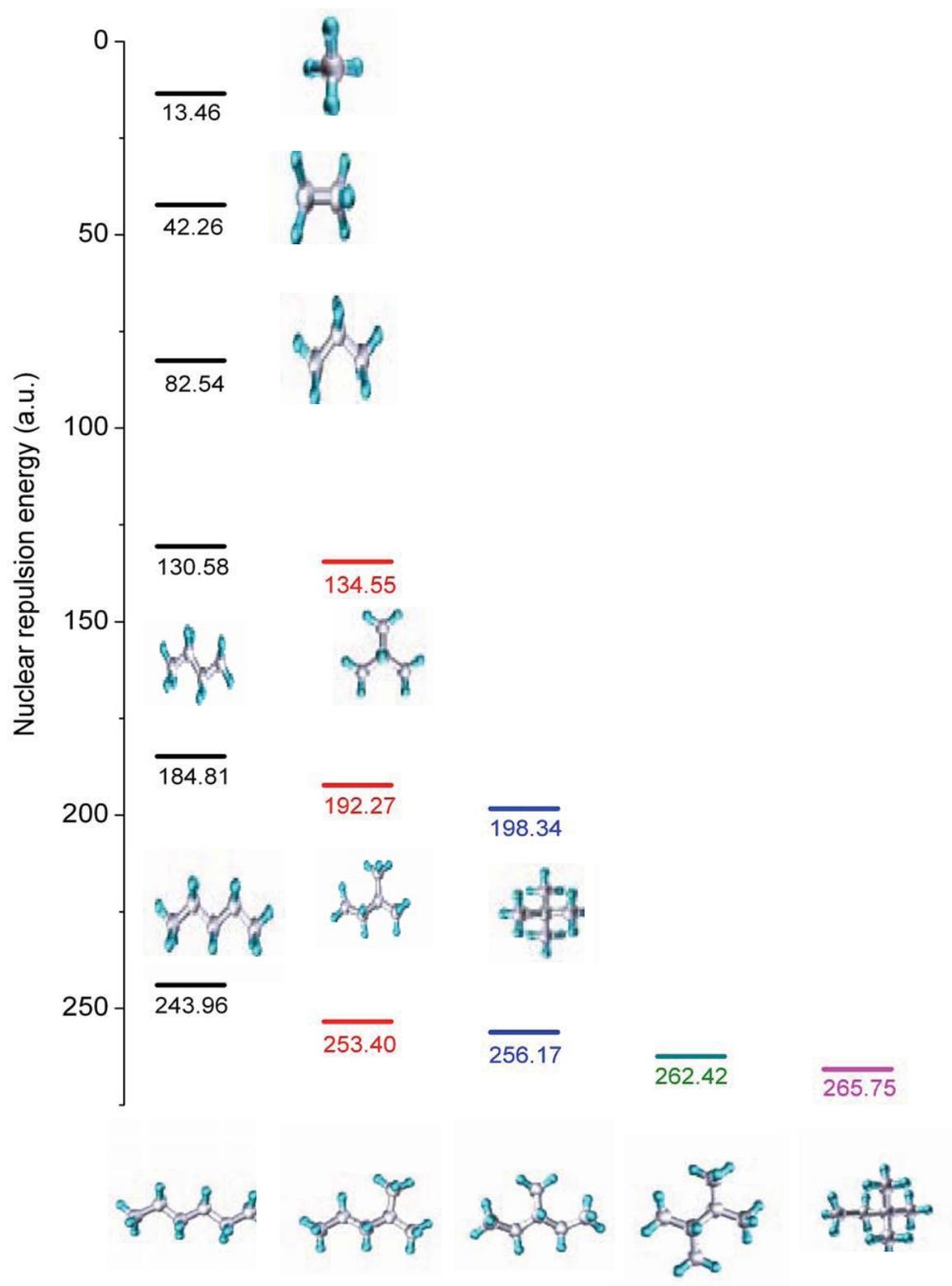

**Figure 1.**

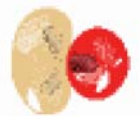

methane (1t$_2$)

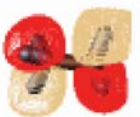

ethane (1e$_g$)

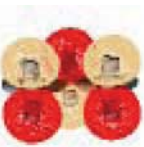

propane (2b$_1$)

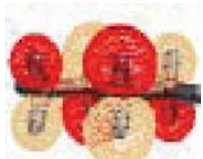

n-butane (2b$_g$)

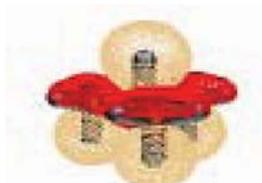

isobutane (6a$_1$)

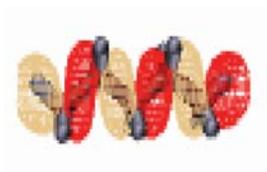

n-pentane (7b$_2$)

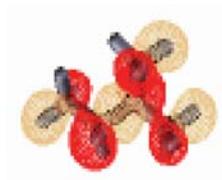

isopentane (21a)

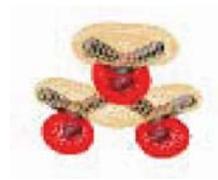

neopentane (4t$_2$)

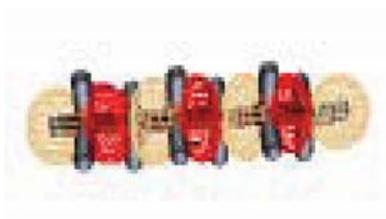

n-hexane (10a$_g$)

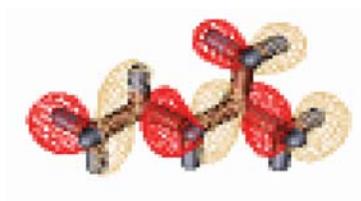

isohexane (25a)

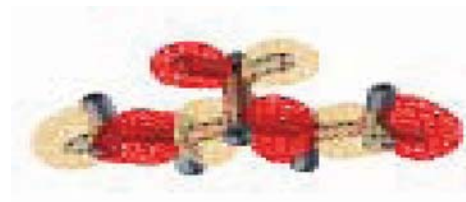

3-methyl pentane (10a")

**Figure 4**.

12